\documentclass[a4paper, titlepage, 10pt, usenames, dvipsnames, svgnames, table]{article}

\usepackage{geometry}

\usepackage{graphicx}
\usepackage[utf8]{inputenc}

\usepackage{color}
\newcommand{\revchange}[1]{#1}

\usepackage{siunitx}
\DeclareSIUnit\kgperL{\kilo\gram\per\liter}
\DeclareSIUnit\waveno{\per\centi\metre}
\DeclareSIUnit\invcm{\per\centi\metre}
\DeclareSIUnit\diffconst{\square\centi\metre\per\second}

\usepackage[version=3]{mhchem}

\usepackage{caption}
\usepackage{subcaption}

\usepackage{multicol}

\usepackage{verbatim}

\newcommand{\fref}[1]{Fig.~\ref{#1}}

\newcommand{\eref}[1]{Eq.~\ref{#1}}

\usepackage[hyperref=false,
            url=false,
            isbn=false,
            doi=false,
            eprint=false,
            sorting=none,
            firstinits=true,        
            maxbibnames=99,
            backend=biber]{biblatex}
\AtEveryBibitem
{
    \clearfield{number}
    \clearfield{issue}
    \clearfield{eid}
    \clearfield{note}
}
\DefineBibliographyExtras{english}{}
\DefineBibliographyExtras{english}{}
\bibliography{literature}
\renewbibmacro{in:}{}

\usepackage{authblk}

\usepackage{bm}
\renewcommand{\vec}[1]{\boldsymbol{\mathbf{#1}}}
\newcommand{\mat}[1]{\mathbf{#1}}

\makeatletter
\renewcommand*{\@fnsymbol}[1]{\ifcase#1\or*\else\@arabic{\numexpr#1-1\relax}\fi}
\makeatother

\begin{document}

\title{Spectroscopy from Machine Learning by Accurately Representing the Atomic Polar Tensor}

\author[1]{Philipp Schienbein\thanks{email: p.schienbein@ucl.ac.uk}}
\affil[1]{Department of Physics and Astronomy and Thomas Young Centre, University College London, London, WC1E 6BT, United Kingdom}
\date{ }

\maketitle

\begin{abstract}
    Vibrational spectroscopy 
    is a 
    key technique to elucidate microscopic structure and dynamics.
    Without the aid of theoretical approaches, it is however, often difficult to understand such spectra at a microscopic level.
    Ab initio molecular dynamics have repeatedly proved to be suitable for this purpose, however, the computational cost can be daunting.
    Here, the E(3)-equivariant neural network \texttt{e3nn} is used to fit the atomic polar tensor of liquid water
    \textit{a posteriori} 
    on top of existing molecular dynamics simulations. 
    Notably, the 
    \revchange{introduced methodology}
    is general and thus transferable to any other system as well.
    The target property is most fundamental, gives access to the IR spectrum and, more importantly,
    it
    is
    a highly powerful tool 
    to directly assign
    IR spectral features 
    to nuclear motion---a connection which has been pursued 
    in the past but only using severe approximations due to the prohibitive computational cost.
    The herein introduced 
    methodology
    overcomes this bottleneck. 
    To benchmark the machine learning model, the IR spectrum of liquid water is calculated, 
    indeed showing
    excellent agreement with the explicit reference calculation.
    In conclusion, the presented methodology gives a new route to calculate accurate IR spectra from molecular dynamics simulations and will facilitate the understanding of such spectra on a microscopic level.
\end{abstract}

Vibrational spectroscopy, be it IR or Raman spectroscopy, is 
one of the most important techniques to unveil microscopic properties of matter, such as structure determination~\cite{Huth-2011-NatMater} or structural dynamics of water~\cite{Bakker-2010-ChemRev, Perakis-2016-ChemRev}.
It can also be used to time-dependently monitor dynamical processes, e.g.\ intramolecular couplings in peptides~\cite{Kolano-2006-Nature}, or proton transfer mechanisms~\cite{Wolke-2016-Science}. 
Attenuated total reflection~(ATR) IR or sum-frequency generation~(SFG)~\cite{Shen-2006-ChemRev} can selectively probe molecules at interfaces, such as metal/liquid water interfaces~\cite{Nayak-2016-PCCP, Gardner-2019-PCCP} or water/air interfaces~\cite{Kusaka-2021-NatChem}.
Using THz spectroscopy the chemical environment of molecules can directly be probed and it elucidated solvation dynamics in aqueous solutions from ambient~\cite{Heyden-2010-PNAS, Schienbein-2017-JPCL} to extreme thermodynamic conditions, such as high-pressures~\cite{Imoto-2016-ANIE}, supercritical phases~\cite{Schienbein-2020-ANIE}, or confined systems~\cite{RuizBarragan-2022-PCCP}.

Because of the broad applicability and power of vibrational spectroscopy 
there is a demand to calculate accurate vibrational spectra from ab initio techniques which are predictive and also aid to understand experiments at the molecular level.
When it comes to spectroscopy of condensed phase systems at finite temperature, ab initio molecular dynamics~(AIMD)~\cite{Marx-Book-2009}, where the electronic structure is calculated on-the-fly at every timestep, is the prime technique
for several reasons.
First, the potential energy surface can be accurately represented---clearly depending on the underlying electronic structure theory 
setup employed to drive the MD.
Second, as the electronic structure is available at every time step, charge transfer and polarization effects are naturally included and 
dipole moments can directly be obtained, e.g.\ from maximally localized Wannier functions~\cite{Marx-Book-2009} or from the electron density~\cite{Thomas-2015-PCCP}.
Third, anharmonic effects are also naturally taken into account which can break the standard normal mode analysis~\cite{Wilson-1955-MolecularVibrations} if they are too large.
Notably,
such large anharmonicities can even be present for
single molecules (e.g.\ peptides) in the gas-phase~\cite{Galimberti-2019-FaradayDiscuss,Gaigeot-2021-SpectrochimActaA}, where AIMD simulations are required 
because the normal mode analysis fails to correctly reproduce the measured spectra.

From a MD trajectory, the frequency dependent Beer-Lambert absorption coefficient of IR spectroscopy
\begin{equation}
    \alpha(\omega) = \frac{\pi\beta\omega^2}{3Vc\epsilon_0 n(\omega)}
        \frac{1}{2\pi}
        \int_{-\infty}^\infty dt \, e^{-i\omega t} \left< \vec{M}(0) \vec{M}(t) \right>
    \label{eq:totspec} 
\end{equation}
can be calculated from the time auto correlation function of the total dipole moment vector $\vec{M}(t)$ of the simulation box (see e.g.\ \cite{McQuarrie2000}), where
$\beta = 1/k_\text{B}T$, 
$k_\text{B}$ is the Boltzmann constant, 
$T$ is the temperature, 
$V$ is the volume of the periodic simulation box, 
$c$ is the speed of light in vacuum,
$\epsilon_0$ is the dielectric constant, 
and
$n(\omega)$ is the frequency dependent refractive index.
Note that the so-called ``quantum correction factor''~\cite{Ramirez-2004-JCP} has already been included. 
The main disadvantage of this technique 
is that 
the MD simulations need to be quite long, 
also
to reduce the statistical noise to a minimum.
Clearly, this is 
a problem for AIMD  which can 
be highly demanding computationally, especially if more expensive techniques, such as hybrid DFT, are used to drive the MD.

Over the last decades, machine learning~(ML) approaches
have been introduced with the aim to accelerate AIMD simulations.
Therein, the expensive electronic structure calculations are replaced with a cheaper machine learning model, while retaining the same accuracy~\cite{Behler-2007-PRL,Bartok-2010-PRL,Schuett-2017-NatCommun, Behler-2021-ChemRev}.
Arguably, ML techniques have repeatedly proved to reliably 
represent
the potential energy surface from explicit electronic structure calculations at a fraction of the cost.
One apparent problem of these pioneering ML techniques usually is that only the potential energy surface is trained (which is generally sufficient to run MD simulations), but all information on the electronic structure itself is lost.
Therefore, total dipole moments at the quality of the underlying electronic structure theory cannot be obtained along the ``MLMD'' (machine learning molecular dynamics) simulation.
One way to circumvent this issue was to extract single snapshots from the MLMD trajectory and explicitly calculate the electronic structure for those snapshots again, e.g.\ to calculate polarizability tensors for Raman spectra~\cite{Morawietz-2018-JPCL}.
However, formally, time correlation functions (as in e.g.\ \eref{eq:totspec} for IR spectra), require the electronic structure at each time step; or at least frequently enough such that all 
vibrations present in the system are correctly sampled.
Note, that the sampling theorem can be employed to determine how frequently time-dependent data needs to be provided~\cite{Shannon-1949-IEEE}, however, the fastest vibration needs to be known.
For example, in case of liquid water, the fastest vibration is the \ce{O-H} stretch at roughly \SI{3500}{\invcm}. 
According to the sampling theorem, data needs to be provided at least every roughly 4.5~fs to correctly sample this vibration.
This introduces a huge bottleneck for MLMD simulations, if a significant amount of configurations needs to be explicitly recalculated anyways to get exact vibrational spectra.

In recent years, training atomic or molecular properties using ML 
has been an extremely active field, 
\revchange{and the calculation of vibrational spectra by ML is no exception.}
\revchange{%
    In the following, some key methodological ideas are summarized how the computation of vibrational spectra can be accelerated by ML.
    Since the approach introduced herein aims to calculate vibrational spectra from MD simulations, i.e.\ via time correlation functions, the following discussion is restricted to accelerating these methods only.
    Notably, ML approaches have been used previously to accelerate complementary approaches, too, e.g.\ the normal mode analysis or vibrational Hamiltonians.
    ML approaches have also been used for the reverse ``Spec-to-Struc'' process, where a given spectrum is used to gain information on the underlying structure.
    The interested reader is referred to Ref.~\cite{Han-2022-JPCA} for a detailed review on ML in the context of these methods 
    and
    on applications of ML in the context of vibrational spectroscopy in general.
}%

Partial atomic charges have been introduced in third generation NNPs~\cite{Behler-2021-ChemRev}, but with the main purpose to 
include long-range interactions.
Such trained atomic partial charges could potentially also be used as an output parameter to calculate dipole moments along a MLMD trajectory.
An apparent problem of partial charges in general is, that they are no physical observables.
As such, their magnitude depends on the chosen partitioning scheme employed, e.g.\ 
Mulliken~\cite{Mulliken-1955-JCP}, Hirshfeld~\cite{Hirshfeld-1977-TheoretChimActa}, or Bader~\cite{Bader-1985-AccChemRes} (incomplete list).
It could be shown that different partitioning schemes can yield very different results~\cite{Wiberg-1993-JComputChem, Sifain-2018-JPCL, Han-2021-JCTC}.
Moreover, choosing an unsuitable scheme for a given problem can lead to wrong molecular dipole moments~\cite{Han-2021-JCTC} and can yield unphysical atomic charges~\cite{Ahart-2022-JCTC}.
Clearly, these caveats potentially also affect the quality of the IR spectrum calculated from such partial atomic charges.

A complementary approach is to train partial charges such, that molecular dipole moments are reproduced correctly~\cite{Gastegger-2017-ChemSci, Sifain-2018-JPCL} or to train the positions of Wannier centers%
\cite{Gao-2022-NatCommun, Cools-2022-JCTC}.
Molecular dipole moment are generally measurable and can thus be validated against experiments. 
In case of gas-phase systems (without periodic boundary conditions), the total dipole moment vector has recently been trained as a whole to predict IR spectra of 
protonated water clusters~\cite{Beckmann-2022-JCTC} as well as of an ethanol and an aspirin molecule~\cite{Schuett-2021-ProcMachineLearningRes}.
For small single molecules, also the polarizability tensor has been trained~\cite{Wilkins-2019-PNAS,Schuett-2021-ProcMachineLearningRes} which can then be used to calculate Raman spectra of these molecules.
Finally, 
even the full electron density of single molecules has been trained~\cite{Grisafi-2018-ACSCentSci, 
Unke-2021-AdvNeuralInfProcessingSys} as well as transition dipole moments to excited states~\cite{Westermayr-2020-JCP} which allow the calculation of UV/Vis spectra.

In condensed phase systems with periodic boundary conditions,
the total dipole moment is, however, multivalued~\cite{Spaldin-2012-JSolidStateChem} which can possibly lead to ambiguities in the training and prediction process.
This problem can be circumvented by considering molecular dipole moments instead.
The total dipole moment vector required to calculate the IR spectrum according to \eref{eq:totspec} is then the sum of all molecular dipole moment vectors in the system.
Molecular dipole moment vectors 
have been trained
directly by symmetry-adapted Gaussian process regression to accurately calculate the IR spectrum of liquid ambient water~\cite{Kapil-2020-JCP}.
Using the same approach, 
it was shown that 
the molecular polarizability tensor can also directly be trained which 
enables one 
to calculate machine learned Raman~\cite{Raimbault-2019-NewJPhys,Kapil-2020-JCP, Shepherd-2021-JPCL} and SFG~\cite{Shepherd-2021-JPCL} spectra from molecular dynamics simulations.

In this work, I introduce a machine learning model to train the atomic polar tensor~(APT)~\cite{Person-1974-JCP} which is then utilized to accurately calculate an IR spectrum.
\revchange{
    The APT is a proper physical observable which does not rely on any charge partitioning scheme, or any definition of molecules or molecular reference frames.
}%
Its definition is therefore also perfectly valid when covalent bonds are broken during a MD simulation and the molecular composition changes, e.g.\ during proton transfer in water.
Conceptually, the herein introduced APT neural network (APTNN) is therefore transferable to any system with or without periodic boundary conditions.
It is noted in passing that nuclear quantum effects are essential to describe e.g.\ proton transfer in water correctly
which are not considered in this work.
However, the introduced APTNN is readily applicable to path integral trajectories since the APT centroid can straightforwardly be computed.

Importantly,
the APT itself
is a highly relevant property for spectroscopy, because it can be utilized to assign specific atomic motion to spectral features.
As it will be laid out in the following, 
the APT represents nothing else than the definition of the IR selection rule.
As such, any 
velocity spectrum (e.g.\ the vibrational density of states) 
can be promoted to a proper IR spectrum by weighting with the respective APTs.
This has been done several times in the past, e.g.\ to decompose the IR spectrum of liquid water into translational, rotational, and vibrational contributions~\cite{Imoto-2019-JCP}.
IR spectra of peptides have also been dissected in terms of atomic velocities in the past~\cite{Jaeqx-2014-ANIE}, even using sophisticated decompositions based on graph theory to 
understand the origin of
low-frequency backbone vibrations~\cite{Galimberti-2019-FaradayDiscuss}.
Moreover, it has also been used to calculate SFG spectra~\cite{Khatib-2017-JPCL}.
None of these works has yet utilized the full power of the APT simply due to the enormous computational cost:
A single APT requires six additional single point calculations to obtain the necessary finite differences
\revchange{
    (if the APT is calculated numerically)
}%
, see below.
As the result, severe approximations have been used so far, such as parametrizing the APT~\cite{Khatib-2017-JPCL}, using the instantaneous normal mode~(INM) approximation~\cite{Imoto-2019-JCP}, or calculating the APT not at every time step, under the approximation that it does not change much as a function of time~\cite{Galimberti-2017-JCTC}.
These approximations clearly counteract its potential power:
The INM approximation typically introduces imaginary frequencies (similar to the standard normal mode analysis) which need to be dealt with in some ad hoc way.
Similarly, calculating the APT not frequently enough can induce spurious signals in the IR spectrum~\cite{Galimberti-2017-JCTC}.
Having a machine learning approach available to specifically predict the APT at each time step is therefore highly beneficial for all above mentioned problems.
To the best of my knowledge such a ML model does not exist yet.

\revchange{
    The derivation of the APT has already been presented in the literature, see e.g.\ Refs.~\cite{Galimberti-2017-JCTC, Khatib-2017-JPCL}.
    To set the stage, its derivation is, however, summarized here.
}%
First,
a Fourier transform identity is used on \eref{eq:totspec}, such that the equation can be rewritten
\begin{equation}
    \alpha(\omega) = \frac{\pi\beta}{3Vc\epsilon_0 n(\omega)}
        \frac{1}{2\pi}
        \int_{-\infty}^\infty dt \, e^{-i\omega t} \left< \dot{\vec{M}}(0) \dot{\vec{M}}(t) \right> \quad ,
        \label{eq:totspec-dotM}
\end{equation}
where $\dot{\vec{M}}(t)$ is the time derivative of the total dipole moment.
In an effort to express the dipolar velocity as a function of atomic velocities, the chain rule can be applied to express the $\xi$-th component of 
$\dot{\vec{M}}(t)$
\begin{equation}
    \dot{M}_\xi(t) = \sum_i^{N_\text{nuc}} \sum_\zeta 
    \frac{\partial M_\xi(t)}{\partial r_{i\zeta}(t)} 
    \cdot
    \frac{\partial r_{i\zeta}(t)}{\partial t}
    \, ,
    \label{eq:apt-sum}
\end{equation}
where $\zeta$ and $\xi$ represent the three cartesian coordinates.
In matrix notation, this equation can be rewritten in a more compact way,
\begin{equation}
    \dot{\vec{M}}(t) = \sum_i \mat{P}_i(t) \cdot \vec{v}_i(t) \quad, 
    \label{eq:totdip-from-apt}
\end{equation}
where $\vec{v}_i(t)$ is the velocity of the $i$-th nuclei and 
\begin{equation}
    \mat{P}_i(t) 
    = \left(\nabla_i \otimes \vec{M}(t)\right)^\text{T}
    \label{eq:apt}
\end{equation}
is the 
APT 
of atom $i$.
The atom velocities $\vec{v}_i(t)$ are readily available along any MD trajectory.
As shown in \eref{eq:apt-sum}, the APT is the spatial derivative of the total dipole moment vector, which is indeed nothing else than the IR selection rule.
Any IR spectrum can therefore be expressed based on atomic velocities weighted by the corresponding APTs.
The atomic velocities in turn can intuitively be dissected \textit{a la carte} using classical mechanics.
The herein introduced APTNN is meant to be trained on existing MD trajectories (AIMD, MLMD).
After the training, the APTs of each 
sampled configuration
can be predicted and the time correlation function can be sampled.
\revchange{
Note, that IR spectra have usually been computed using the autocorrelation function depicted in \eref{eq:totspec}
from explicit AIMD simulations where electric dipole moments are straightforwardly available.
Training the APT therefore only becomes relevant either if the APT is used for the spectral analysis itself, see e.g.\ Refs.~\cite{Khatib-2017-JPCL, Galimberti-2017-JCTC, Imoto-2019-JCP, Galimberti-2019-FaradayDiscuss}, or if
sampling is performed by MLMD simulations where the total dipole moment is usually not naturally available.
The latter is especially important when longer timescales are required which exceed the ones accessible by explicit AIMD or when more expensive electronic structure calculations, e.g.\ hybrid DFT or beyond, are applied.
}%

The APT of atom $i$ is the spatial derivative of the total dipole moment with respect to a displacement of that atom (\eref{eq:apt}). 
It is thus a 3x3 tensor which is invariant with respect to translations, but equivariant with respect to rotations.
This means that if the atomic coordinates are translated in space, the corresponding APTs do not change.
However, if the set of atomic coordinates rotates in space, the APTs also rotate accordingly.
\revchange{
Here, 
the derivative is evaluated numerically from central (or ``two-sided'') finite differences as it has frequently been done before, see e.g.~\cite{Khatib-2017-JPCL, Galimberti-2019-FaradayDiscuss, Imoto-2019-JCP}, using a displacement of \SI{0.01}{\angstrom}.
}
Due to the finite differences, 6 additional single point calculations are necessary to calculate the APT of a single atom. 
Previously, it could already be shown that the chosen displacement is small enough for liquid water~\cite{Imoto-2019-JCP}. 
Indeed, I also computed the APT explicitly with a displacement of \SI{0.04}{\angstrom} and did not find any significant difference.
It might, however, not be sufficiently small enough for other systems.
\revchange{
    Notably, it is also possible to calculate an APT analytically using Density Functional Perturbation Theory~(DFPT) as implemented in \texttt{CP2k}~\cite{Ditler-2021-JCP}.
    This way, only one electronic structure calculation is required per atom per cartesian coordinate to calculate the APT. 
    The total number of required electronic structure calculations would thus be reduced compared to using numerical derivatives.
    More importantly, the displacement for the numerical derivative is a convergence parameter which could be omitted completely when using analytical derivatives.
    However, the herein employed numerical derivatives render the presented methodology general, such that it
    can directly be applied even if analytical APTs are not available, e.g.\ when other codes or electronic structure methods beyond DFT are used. 
}

An apparent problem when training an APT (see \eref{eq:apt}) is that it is an equivariant property as already mentioned, while most machine learning models can only infer invariant properties.
Recently, the 
\texttt{e3nn} framework has been introduced~\cite{e3nn} 
for PyTorch
which can be used to
train E(3)-equivariant graph Neural Networks and thus enables one to
infer also equivariant properties by a machine learning model.
Its introduction also caused a huge boom in the field and numerous works have been published, where equivariant properties have been modeled.
For example, the \textit{NequIP} package was recently developed~\cite{Batzner-2022-NatCommun} to model potential energy surfaces using equivariant message passing.
Moreover, it was also used recently to train the electron density as a whole for gas-phase molecules~\cite{Unke-2021-AdvNeuralInfProcessingSys}.
Besides of such equivariant message passing Neural Networks, also Gaussian process regression can be used to create equivariant machine learning models~\cite{Grisafi-2018-PRL}.
This approach was utilized recently to train molecular dipole moment vectors and polarizability tensors~\cite{Raimbault-2019-NewJPhys, Kapil-2020-JCP, Shepherd-2021-JPCL}.

Here, I now use the 
so-called \textit{SimpleNetwork} (v2106) from the \texttt{e3nn} 
toolkit 
to create an APTNN for liquid ambient water which is capable to predict the APTs of all atoms at a given MD snapshot.
The predicted APTs are then used to calculate the IR spectrum 
\revchange{via \eref{eq:totspec-dotM} and \eref{eq:totdip-from-apt}}.
Although the IR spectrum of liquid ambient water 
has already been machine learned by training
molecular dipole moments~\cite{Kapil-2020-JCP,Shepherd-2021-JPCL}, even explicitly considering nuclear quantum effects, it is merely done here to demonstrate the applicability of the herein introduced technique. 
The introduced APTNN \revchange{methodology and the training protocol} is generally transferable to any other system, with and without periodic boundary conditions.
As the underlying electronic structure theory, I opt to use the
RPBE functional~\cite{Hammer-1999-PRB} supplemented by D3 dispersion corrections~\cite{Grimme-2010-JCP}.
It could be shown repeatedly in the past, that this electronic structure model reproduces fluid water excellently, even far away from ambient conditions~\cite{Imoto-2015-PCCP,Schienbein-2020-PCCP, Schienbein-2020-ANIE, Gross-2022-ChemRev}.
Previously, liquid ambient water has been simulated 
\cite{Imoto-2015-PCCP} and these data (16 trajectories, 20~ps each using a timestep of 1~fs) are used in this work.
\revchange{
    Note that the APTNN meant to be trained on top of existing MD trajectories. 
    This means that these underlying MD trajectories are entirely responsible for sufficiently sampling 
    the configuration space.
}%
As a starting point, I randomly selected 10 statistically independent configurations from the available AIMD trajectories and calculated the APT.
The calculation setup is exactly the same as before~\cite{Imoto-2015-PCCP} and I refer to that reference for an elaborate description.
All electronic structure calculations have been performed using version 8.0 of the \texttt{CP2k} program package~\cite{Kuehne-2020-JCP} and the \texttt{Quickstep module}~\cite{VandeVondele-2005-ComputPhysCommun}.
Each configuration contains 384 atoms and therefore 2304 single point calculations are required. 
This is undoubtedly a substantial computational commitment, however, this way 384 APTs are obtained from a single MD snapshot, which is quite a lot of training data which will become apparent in the following.

Having 10 MD snapshots with explicitly calculated APTs for all atoms available, an APTNN is trained by randomly selecting 9 (90~\%) configurations for its training set.
The remaining configuration is used to validate how well the model generalizes during the training.
\revchange{
    Recall, that the APT is calculated for each atom in each of the 10 snapshots and a single configuration contains 128 water molecules, i.e.\ 384 atoms.
    The training and test set therefore consist of 3456 and 384 APTs, respectively.
    I use a \texttt{e3nn} \textit{SimpleNetwork} (v2106) model consisting of two message passing layers and a technical feature configuration ``\texttt{20x0o+20x0e+20x1o+20x1e+20x2o+20x2e}'', describing the feature set of each atom.
    The latter string encodes that each atom is represented by a feature vector containing 20 scalars (tensor rank 0), 20 vectors (tensor rank 1), and 20 tensors of rank 2 with even (``e'') and odd (``o'') parity each.
    In a nutshell, the input geometry is translated into a graph representation, where each atom is represented by a node and each interatomic connection is represented by an edge.
    Through the message passing layers, the feature vector of each node (atom) is iteratively refined, taking the graph edges (interatomic distance vectors) and the feature vectors of all neighboring nodes (atoms) into account.
    Thereby the feature vectors are optimized, such that they contain a unique representation of the environment of each atom.
    After the message passing phase, the feature vectors are then used to predict an APT in a given configuration.
    The interested reader is referred to Ref.~\cite{Batzner-2022-NatCommun} for an elaborate discussionon how the feature vectors are iteratively refined in the \texttt{e3nn} framework.
    Here, I use a radial cutoff of \SI{6}{\angstrom} to create the graph from the input geometry. 
    This means that only edges between two nodes are added to the graph, if the interatomic distance is smaller than \SI{6}{\angstrom}.
    The radial cutoff is mainly implemented for computational efficiency.
    It ensures that each atom has approximately the same number of neighbors in the graph representation, irrespective of the total system size. 
    This effectively reduces the computational cost, since the number of graph edges (connections between atoms) is limited to the local environment of each atom only.
    Moreover, the computational cost then scales only linearly with the total number of atoms in the system~\cite{Batzner-2022-NatCommun}.
    The radial cutoff clearly is a convergence parameter which needs to be tested to be large enough.
    Here, the actual cutoff value of \SI{6}{\angstrom} was chosen because it was proved previously that it is large enough to correctly reproduce the potential energy surface of liquid ambient water using High-Dimensional NNPs~\cite{Morawietz-2016-PNAS} as well as graph Neural Networks~\cite{Batzner-2022-NatCommun}.
    Note, that I will also show in the following that the cutoff is large enough to train the APT and to accurately reproduce the IR spectrum of liquid ambient water.
}

\revchange{
}%
The model has been trained using the Adam optimizer~\cite{Adam-optimizer}
implemented in pyTorch~\cite{pytorch}.
\revchange{
    The Adam optimizer is one of the most frequently used optimizers in pytorch and has been successfully applied when training \texttt{e3nn} based models in the past, see e.g.\ Ref.~\cite{Batzner-2022-NatCommun}.
}%
An initial learning rate of 0.01 is used which is automatically reduced by a factor of 0.1, when the loss of the validation data set does not decrease further over the last 10 training epochs.
The hyperparameters have been chosen manually based on the validation set performance during the training and have then been fixed.
The overall performance of the trained model was finally evaluated on an unrelated test set and on the predicted IR spectrum compared with available reference and experimental data, see below.

\begin{figure}
    \begin{subfigure}{0.49\textwidth}
        \centering
        \includegraphics[width=\textwidth]{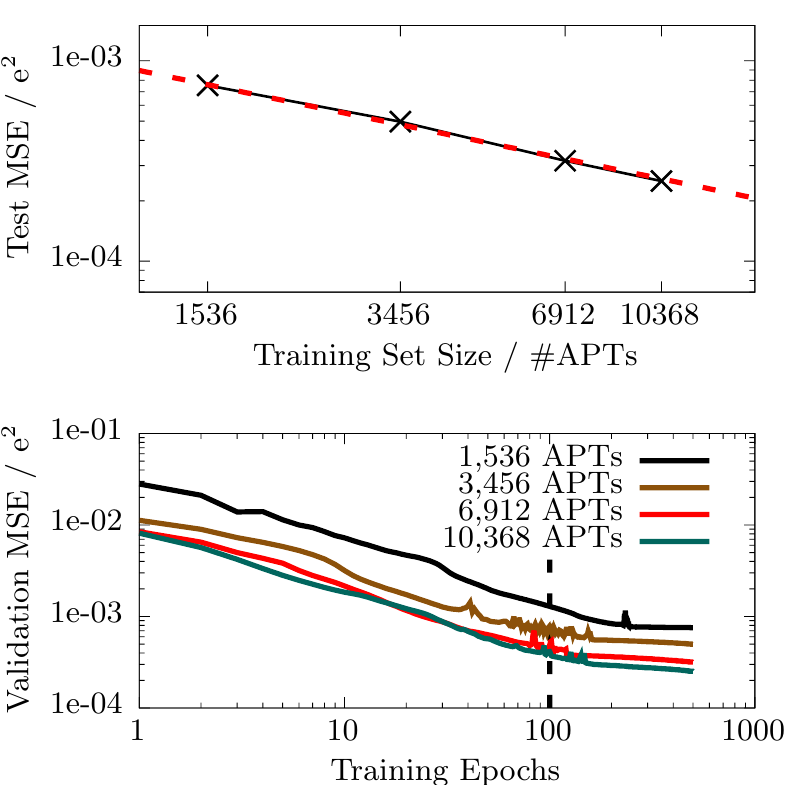}
        \caption{}
        \label{fig:learning-curve}
    \end{subfigure}
    \begin{subfigure}{0.49\textwidth}
        \centering
        \includegraphics[width=\textwidth]{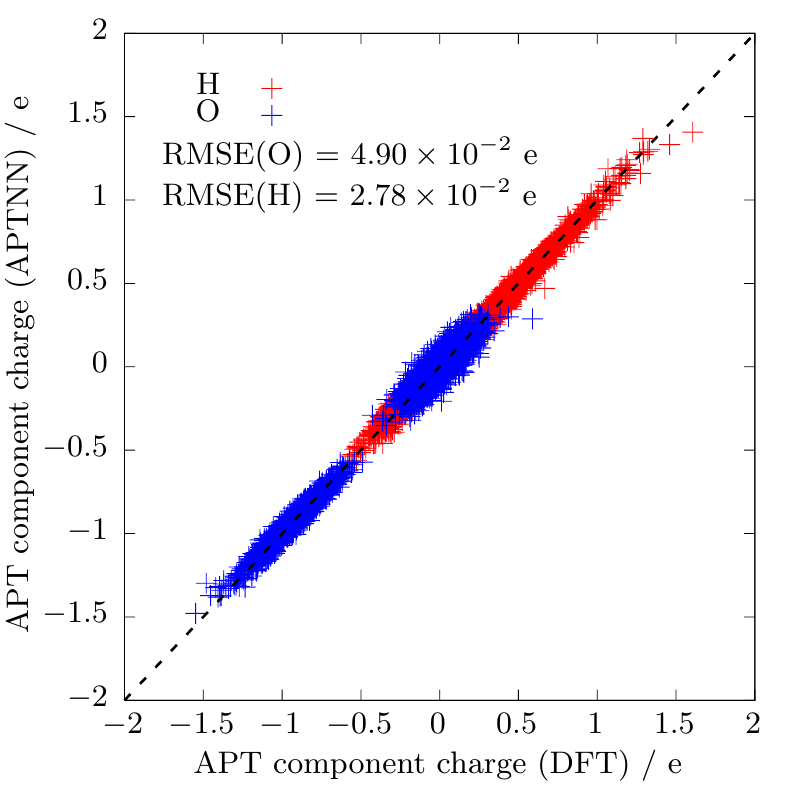}
        \caption{}
        \label{fig:rmse-scatter}
    \end{subfigure}
    \caption{
        \revchange{
            (a)
            Test error evaluated on an unknown test set containing 3840 APTs stemming from 10 randomly sampled MD snapshots as a function of training data set size (learning curve, top panel).
            Mind the log-log scale of the plot and that the expected linear relation (indicated by the red dashed line) is recovered.
            In the bottom panel, the MSE on the respective validation set is shown as a function of training epoch and training data set size, i.e.\ 4 (black), 9 (brown), 18 (red), and 27 (green) MD snapshots, corresponding to 1536, 3456, 6912, and 10368 APTs, respectively.
            (b)
            Performance of the APTNN trained on 3840 APTs (9 MD snapshots) on an unknown test set containing 3840 APTs stemming from 10 randomly sampled MD snapshots.
            The figure compares all components of the APT matrix individually for O (blue) and H (red) atoms. 
            Note that the test set used to benchmark the APTNN in the top panel of (a) and in (b) is the same.
        }
    }
\end{figure}

The learning curve
of a APTNN  model
as a function of the data set size (4, 9, 18 and 27 training configurations, 
\revchange{
    corresponding to 1536, 3456, 6912, and 10368 APTs, respectively)
    is shown in the top panel of \fref{fig:learning-curve}.
}%
All 
APTNNs
have been trained 
according to the above described training procedure.
The presented \revchange{test} mean squared error~(MSE) is computed 
on 
\revchange{
    an unknown test set containing 3840 APTs in total stemming from 10 randomly sampled MD snapshots.
}
The
test error systematically decreases 
as a function of provided training data set size,
\revchange{
following a linear behavior in the log-log plot as expected.
}%

All data presented in the following are calculated from the 
APTNN trained using 
9 configurations only (solid brown learning curve in \revchange{the bottom panel of} \fref{fig:learning-curve}) after 100 epochs (marked by a vertical dashed line).
To \revchange{benchmark} this model,
I randomly select 10 configurations from the available AIMD simulations of liquid water which have 
not been included in the training set (``test set'').
For those configurations, the APT is explicitly calculated using the electronic structure method as before and compared with the prediction of the APTNN.
The direct component-wise comparison is shown in \fref{fig:rmse-scatter} and the overall component-wise RMSE is \num{3.49e-2}~e, where e is the electron charge.

\begin{figure}
        \centering
        \includegraphics[width=0.5\textwidth]{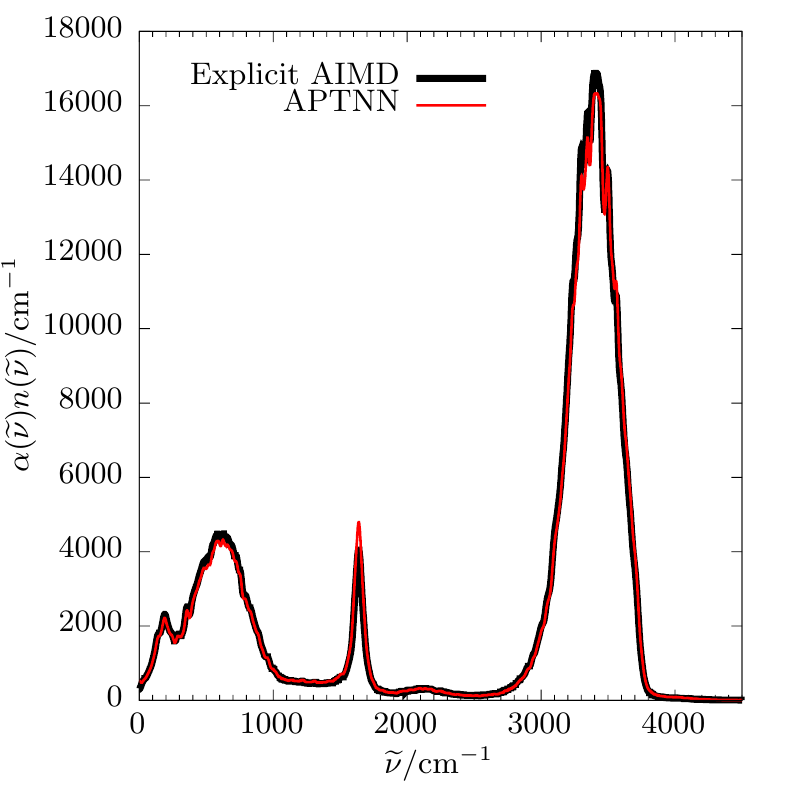}
    \caption{
        In (a) the machine learned IR spectrum 
        of liquid ambient water (red) is compared with the explicit AIMD reference spectrum (black), see text.
    }
    \label{fig:spectrum}
\end{figure}

As a second quality benchmark, I now use the trained APTNN 
to predict the IR spectrum of liquid ambient water.
In practical terms, I take all the available AIMD trajectories of RPBE-D3 liquid ambient water (16 trajectories, 20~ps each using a timestep of 1~fs, see above) 
and predict the APT for each atom at each time step.
Recall that 
this is 
a daunting 
task for explicit electronic structure calculations.
Having all these APTs available, I then calculate the total dipole moment derivative at each time step according to \eref{eq:totdip-from-apt}.
Finally, the machine learned IR spectrum is obtained by \eref{eq:totspec-dotM} and presented in \fref{fig:spectrum}.
The figure also shows 
the reference IR spectrum which has been obtained directly from the explicit AIMD simulations.
Note that this reference IR spectrum of RPBE-D3 water has already 
been published before, see e.g.\ Ref.~\cite{Schienbein-2020-ANIE}.
The comparison shows that the machine learned spectrum reproduces the explicitly calculated reference spectrum exactly in shape and intensity.

Remarkably, this accuracy is reached by only considering 9 randomly selected configurations for training, i.e.\ without even using more sophisticated active learning techniques.
\revchange{
Indeed, the \texttt{NequIP} package which also utilizes the \texttt{e3nn} library to train the potential energy surface also showed a high data efficiency~\cite{Batzner-2022-NatCommun}.
This means that a very good agreement with the reference data could be achieved with providing a comparably small training set.
Interestingly, the data efficiency could directly be traced back to the equivariance of the model, since the performance of the model significantly declined as the equivariance was disabled~\cite{Batzner-2022-NatCommun}.
}
\revchange{This data efficiency}
foreshadows that training an APTNN with even more computationally expensive electronic structure theory will be possible; 
indeed,
potential energy surfaces~\cite{Schran-2020-JCTC, Batzner-2022-NatCommun, Beckmann-2022-JCTC} 
and polarizability tensors~\cite{Wilkins-2019-PNAS}
of coupled cluster accuracy have been successfully trained already.
This could for example be achieved using a committee of APTNNs which straightforwardly enables active learning.
Committees have recently shown to be an efficient tool to obtain a highly accurate model while the training set size is minimized~\cite{Schran-2020-JCP, Schran-2021-PNAS, Schienbein-2022-PCCP}.

The acceleration gained by the APTNN is significant.
To have a fair comparison, a benchmark was conducted on the very same
machine (16 Intel(R) Xeon(R) E5-2640 v3 CPUs, 128 cores in total).
A single MD step including wavefunction extrapolation took about 8~s using the GGA functional RPBE. 
Calculating an APT for a single atom requires six single points and 
therefore amounts to 48~s.
Predicting an APT for a single atom using the APTNN however only takes about 8~ms, giving an acceleration of more than three orders of magnitude.
Note that the acceleration becomes even larger when more expensive electronic structure methods are used, e.g.\ hybrid DFT or beyond.
In that case, the computational time to calculate a single point increases dramatically, while the time needed to predict a single APT from the APTNN remains constant.
Moreover, pyTorch and \texttt{e3nn} are designed to run more efficiently on GPUs than CPUs.
A significant additional acceleration is therefore expected when switching to GPUs, however, this has not been tested yet.

In conclusion, an equivariant neural network has been used to model the atomic polar tensors in liquid ambient water.
From this machine learned model, the IR spectrum was calculated which showed excellent agreement with the explicit reference calculation.
Thereby it was demonstrated that the training of the atomic polar tensors is indeed possible.
\revchange{
The presented methodology is transferable to any other system class as well.
This transferability simply arises by the atomic polar tensor itself:
It reduces an IR spectrum to a very fundamental basis, namely atomic motion (given by the atom velocity) and the dipolar changes caused by this motion (given by the atomic polar tensor), corresponding to the IR selection rule.
It is therefore a fundamental physical property which is rigorously defined for any atom.
Moreover, it does not rely on any definition of molecules or charge partitioning schemes.
This virtue is preserved by the herein introduced model and the latter thus 
formally \emph{must} be able to describe any atomistic system.
Indeed, 
explicit ab initio molecular dynamics simulations showed that vibrational spectra of e.g.\ molecules or clusters~\cite{Galimberti-2017-JCTC, Galimberti-2019-FaradayDiscuss, Ditler-2021-JCP}, solids~\cite{Ditler-2021-JCP}, liquids~\cite{Imoto-2019-JCP}, or solid/liquid interfaces~\cite{Khatib-2017-JPCL} can faithfully be described by the atomic polar tensor~\cite{Gaigeot-2021-SpectrochimActaA}. 
}

\revchange{
    Atomic polar tensors have been used in the past to understand IR spectra at the microscopic level since they allow one to dissect the spectrum in terms of atomic velocities. 
    This feature was utilized several times in the past already~\cite{Galimberti-2017-JCTC, Galimberti-2019-FaradayDiscuss, Gaigeot-2021-SpectrochimActaA, Imoto-2019-JCP}, however, only using severe approximations due to the prohibitive computational cost.
    This computational bottleneck can be overcome with the herein presented model.
    Importantly, 
    it was demonstrated, 
    that the latter 
    does not compromise on accuracy compared to the presented exhaustive explicit ab initio molecular dynamics benchmark.
    Moreover, to achieve that accuracy, 
a surprisingly small training data set was required which foreshadows that it might even be possible to train an APTNN on more expensive electronic structure calculations, such as hybrid DFT or even beyond.
    Machine learned MD simulations have already been performed using hybrid DFT~\cite{Schienbein-2022-PCCP} or even CCSD(T)~\cite{Daru-2022-PRL} and allow one to sample nanoseconds of MD trajectories at that level of theory.
    Such high level MD trajectories can easily be postprocessed by the APTNN model
    to obtain well converged vibrational spectra, which are clearly computationally prohibitive 
    otherwise.
}

\revchange{
Given the generality of the atomic polar tensor and the rather small training data set required to accurately train the herein presented APTNN model, the latter has the potential to significantly contribute towards novel physical findings in various systems, especially where large-scale MD simulations or expensive electronic structure calculations are required.
Notably, the herein employed modology can also be generalized to other 3x3 tensors, such as the polarizability tensor which is required for Raman and SFG spectra.
A general toolkit to automatically and efficiently train an APTNN on top of existing (machine learned) molecular dynamics trajectories is currently being developed.
}

\vspace{2cm}
\noindent
{\Large
\textbf{Acknowledgement}
}
\vspace{0.5cm}

I thank Johannes P.\ Dürholt (Evonik Operations) for fruitful discussions and his critical comments on the manuscript as well as
Tess Smidt (MIT) for her help and
guidance at an early stage of the project.
This work was supported by an individual postdoc grant funded by 
the German National Academy of Sciences Leopoldina under grant number LPDS 2020-05.

\clearpage
\printbibliography

\end{document}